\documentclass[english,preprint,superscriptaddress, amsmath,amssymb]{revtex4-2}

\usepackage{graphicx}  
\usepackage{dcolumn}   
\usepackage{bm}        
\usepackage{amssymb}   
\usepackage{color}
\usepackage{units}
\usepackage{chemformula}
\usepackage{amsmath}
\usepackage{natbib}
\usepackage{esint}
\usepackage{ulem}
\usepackage{float}
\usepackage[english]{babel}
\usepackage[colorlinks=true, citecolor=blue]{hyperref}
\usepackage{xcolor}
\begin{document}

\title{Transition from three- to two-dimensional Ising superconductivity in \textbf{few-layer} \ch{NbSe2} by proximity effect from van der Waals heterostacking}
	
	\author{Prakiran Baidya}
	\affiliation{Department of Physics, Indian Institute of Science, Bangalore 560012, India}
	\author{Divya Sahani}
	\affiliation{Department of Physics, Indian Institute of Science, Bangalore 560012, India}
	\author{Hemanta Kumar Kundu}
	\affiliation{Weizmann Institute of Science, Rehovot 7610001, Israel}
	\author{Simrandeep Kaur}
	\affiliation{Department of Physics, Indian Institute of Science, Bangalore 560012, India}
	\author{Priya Tiwari}
	\affiliation{Department of Physics, Indian Institute of Science, Bangalore 560012, India}
	\author{Vivas Bagwe}
	\author{John Jesudasan}
	\affiliation{Tata Institute of Fundamental Research, Mumbai 400005, India}
	\author{Awadhesh Narayan}
	\affiliation{Solid State and Structural Chemistry unit, Indian Institute of Science, Bangalore 560012, India}
	\author{Pratap Raychaudhuri}
	\affiliation{Tata Institute of Fundamental Research, Mumbai 400005, India}
	\author{Aveek Bid}
	\affiliation{Department of Physics, Indian Institute of Science, Bangalore 560012, India}
	\email{aveekbid@iisc.ac.in}
	

	\begin{abstract}
	We report the experimental observation of Ising superconductivity in 3-dimensional \ch{NbSe2} stacked with single-layer \ch{MoS2}. The angular dependence of the upper critical magnetic field and the temperature dependence of the upper parallel critical field  confirm the appearance of two-dimensional Ising superconductivity in the 3-dimensional \ch{NbSe2} with single-layer \ch{MoS2} overlay. We show that the superconducting phase has strong Ising spin-orbit correlations which make the holes spin non-degenerate. Our observation of Ising superconductivity in heterostructures of few-layer \ch{NbSe2} of thickness $\sim15$~nm with single-layer \ch{MoS2} raises the interesting prospect of  observing topological chiral superconductors with nontrivial Chern numbers in a momentum-space spin-split fermionic system.		
	\end{abstract}
	
	\maketitle

\section{Introduction}
	
An important focus in condensed matter physics is on creating 2-dimensional (2D) topological chiral superconductors (SC) with non-trivial Chern numbers~\cite{lutchyn2018majorana, Frolov2020, doi:10.1146/annurev-conmatphys-030212-184337, Nadj-Perge602}. Naturally occurring topological SC being rare~\cite{Kallin_2012, Kallin_2016}, recent efforts have concentrated on inducing superconductivity in systems with strong spin-orbit coupling (SOC)~\cite{fu2008superconducting}. The fundamental idea is that in a time-reversal-invariant non-centrosymmetric system with strong spin-valley locking, the spin degeneracy (in momentum space) of the fermions gets lifted. In such a system, when the chemical potential of the system lies in between the spin-split valence band maxima, repulsive interactions can lead to topological pairing mechanisms~\cite{triola2016general,hsu2017topological}.  

For a single atomic layer of a system with hexagonal lattice, the measured in-plane critical magnetic field is significantly larger than the Pauli limit. The origin of this enhancement lies in the intrinsic Ising SOI (arising as a consequence of in-plane inversion symmetry breaking) in the layer. This SOC manifests itself as an effective out-of-plane Zeeman-like field, $B_{SO}$ with opposite orientations at $K$ and $K'$ valley. Thus, for a cooper pair in such a system, spin-valley locking leads to a large $B^{||}_{c2}$. Ising superconductivity is thus the most promising candidate   for the observation of topological effects. This phenomena have earlier been observed in ionic-gated 2D-\ch{MoS2}~\cite{lu2015evidence,saito2016superconductivity} and in  single-layer \ch{NbSe2}~\cite{xi2016ising}. 

Single-layer of 2H-\ch{MoS2} (hereafter referred to as SL-\ch{MoS2}), owing to its strong SOC~\cite{zhu2011giant}, holds the promise of forming the base for such exotic SC with tailor-made properties. There have been previous reports of Ising superconductivity induced in ionic-gated 2D-\ch{MoS2}~\cite{lu2015evidence,saito2016superconductivity}. But both these methods suffer from certain inherent drawbacks -- structural changes in the material, rapid degradation of the sample under ionic liquid, the difficulty of fabricating a complex device structure and a superconducting transition temperature much lower than theoretical predictions. Single-layer \ch{NbSe2} degrades very rapidly on exposure to the ambient making it unfeasible as the basis for complex device architectures.  A practical path to overcome this hurdle is by engineering novel heterostructures of \ch{NbSe2} with other materials having properties whose combination can give rise to an atmosphere stable Ising superconducting phase.  
	
Additionally, in all cases, the SL-\ch{MoS2} had been electron-doped to achieve the SC. Note that the strength of SOC (and the resulting spin-valley coupling and spin-splitting of the bands) in single-layer TMD is significant in the valence band. This implies that in order to have spinless fermions, such a system must be hole-doped~\cite{triola2016general,hsu2017topological}. A controllable mechanism of inducing 2-dimensional superconductivity in transition metal dichalcogenide  that can achieve relatively high transition temperatures while maintaining the long-term stability of the device is elusive.
		
In this letter, we report the observation of Ising superconductivity in few-layer \ch{NbSe2}($\sim15$~nm) with an overlayer of single-layer \ch{MoS2}. Having both materials from the TMD family avoids many of the  problems arising from lattice mismatch at the interface~\cite{PhysRevLett.113.067003, PhysRevB.90.085128, PhysRevLett.124.236402, doi:10.1021/acsnano.9b07475}. Through systematic magnetotransport measurements, we establish that system develops an Ising superconductivity that is distinct from that of pristine \ch{NbSe2}. Though Ising superconductivity has been observed in ionic-gated \ch{MoS2}~\cite{lu2015evidence, saito2016superconductivity} and in single-layer \ch{NbSe2}, our approach has distinct advantages over previous attempts in terms of yield, stability of the system, and is more amenable for the realization of complex device structures. 

		\begin{figure}[t]
			\begin{center}
				\includegraphics[width=.75\columnwidth]{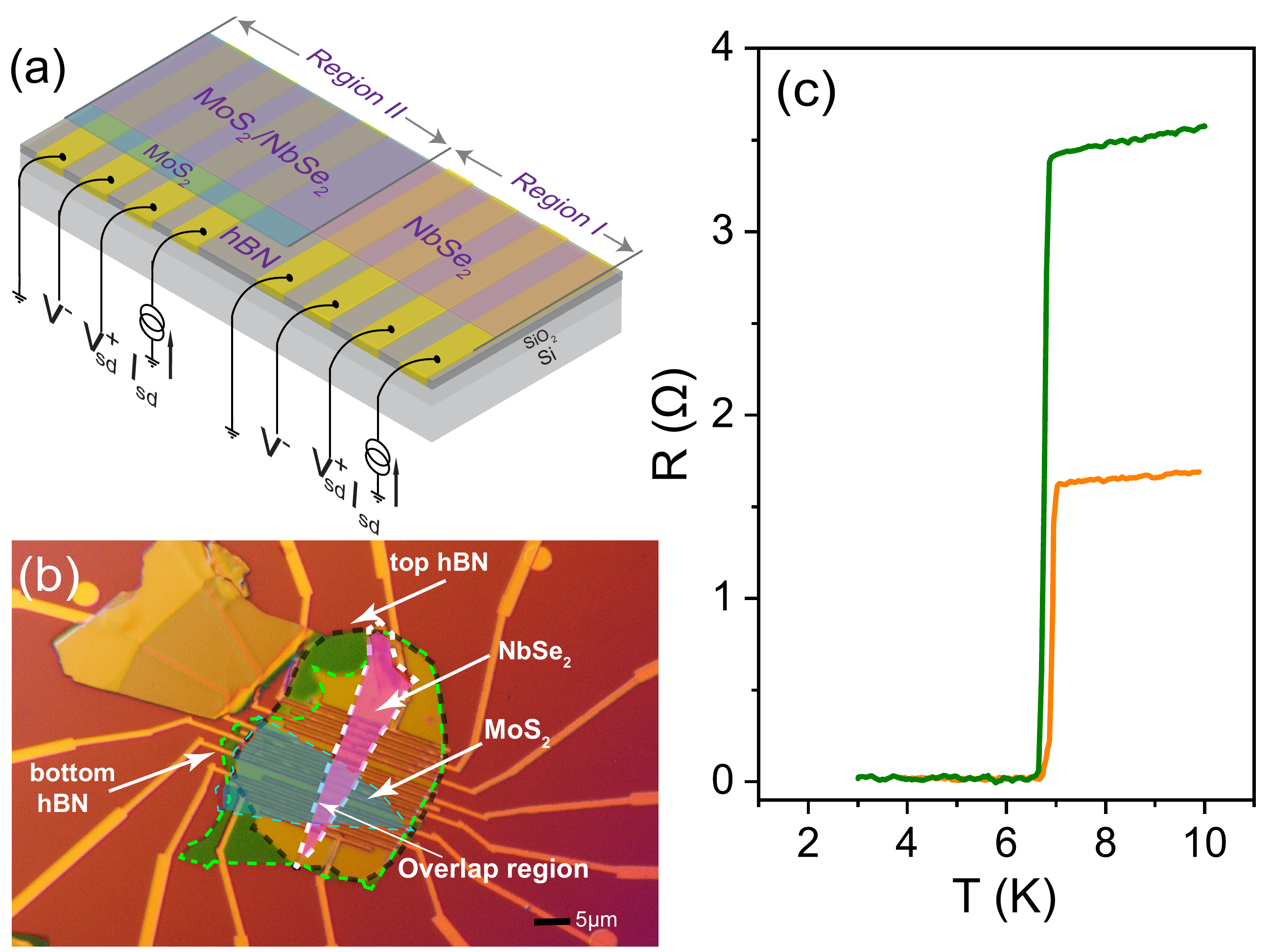}
				\caption{\small {(a) A schematic of the device structure. The  SL-\ch{MoS2}  is transferred on top of gold contact pads etched in hBN. A multi-layer \ch{NbSe2} is transferred  on top such that it lies partially on the \ch{MoS2} (shown in light blue) and partially on the gold probes (shown as an orange rectangle). (b) False color DIC image of a final device. (c) Temperature dependence of the four-probe resistance of the pristine \ch{NbSe2} (orange line) and of the SL-\ch{MoS2}  (olive line) sections of the device.}\label{Fig.1}}   
			\end{center}
		\end{figure}
		
\section{Device fabrication}

The devices were fabricated by transfer of  SL-\ch{MoS2} using the dry transfer technique on gold probes pre-patterned on a thick hBN substrate. This was followed by the transfer of a \ch{NbSe2} flake (thickness $\sim 15$~nm) such it lay  partially on the  SL-\ch{MoS2} and partially on the Au-probes. We refer to the pristine \ch{NbSe2} part of the device as `region-I' and the SL-\ch{MoS2}/\ch{NbSe2} heterostructure as `region-II'. The entire device was capped with a thin flake of hBN (thickness $\sim 30$~nm) to protect the device from environmental degradation (for details of device fabrication, see Appendix~\ref{B}). This device architecture allowed us to study electrical transport simultaneously in both the pure \ch{NbSe2} and in the  SL-\ch{MoS2}/ lying on top of \ch{NbSe2}~(Fig.~\ref{Fig.1}(a)).  Fig.~\ref{Fig.1}(b) is the false-color optical image of a final device showing the SL-\ch{MoS2} (blue area), the \ch{NbSe2}($\sim15$~nm) (pink area, region-I) and the overlap area (region-II).  

\section{Results and Discussions}
		
Electrical transport properties were measured in four-probe configurations using standard low-frequency lock-in detection technique with an excitation current of $1\mu$A in a dilution refrigerator. At room temperature and under zero gate bias, the  SL-\ch{MoS2} capped by \ch{NbSe2} (`region-II') had significantly lower resistance ($\sim 700 \Omega$) as compared to that measured for \ch{MoS2} on \ch{SiO2} or on hBN substrates ($>1 M\Omega$) (data not shown). Fig.~\ref{Fig.1}(c) shows the temperature dependence of the resistance $R$ of the two regions for $T<10$~K -- the resistances of both  region-I and region-II decrease with decreasing temperature  $T$ before becoming smaller than our measurement resolution. 
		
\begin{figure}[t]
		\begin{center}
		\includegraphics[width=.75\columnwidth]{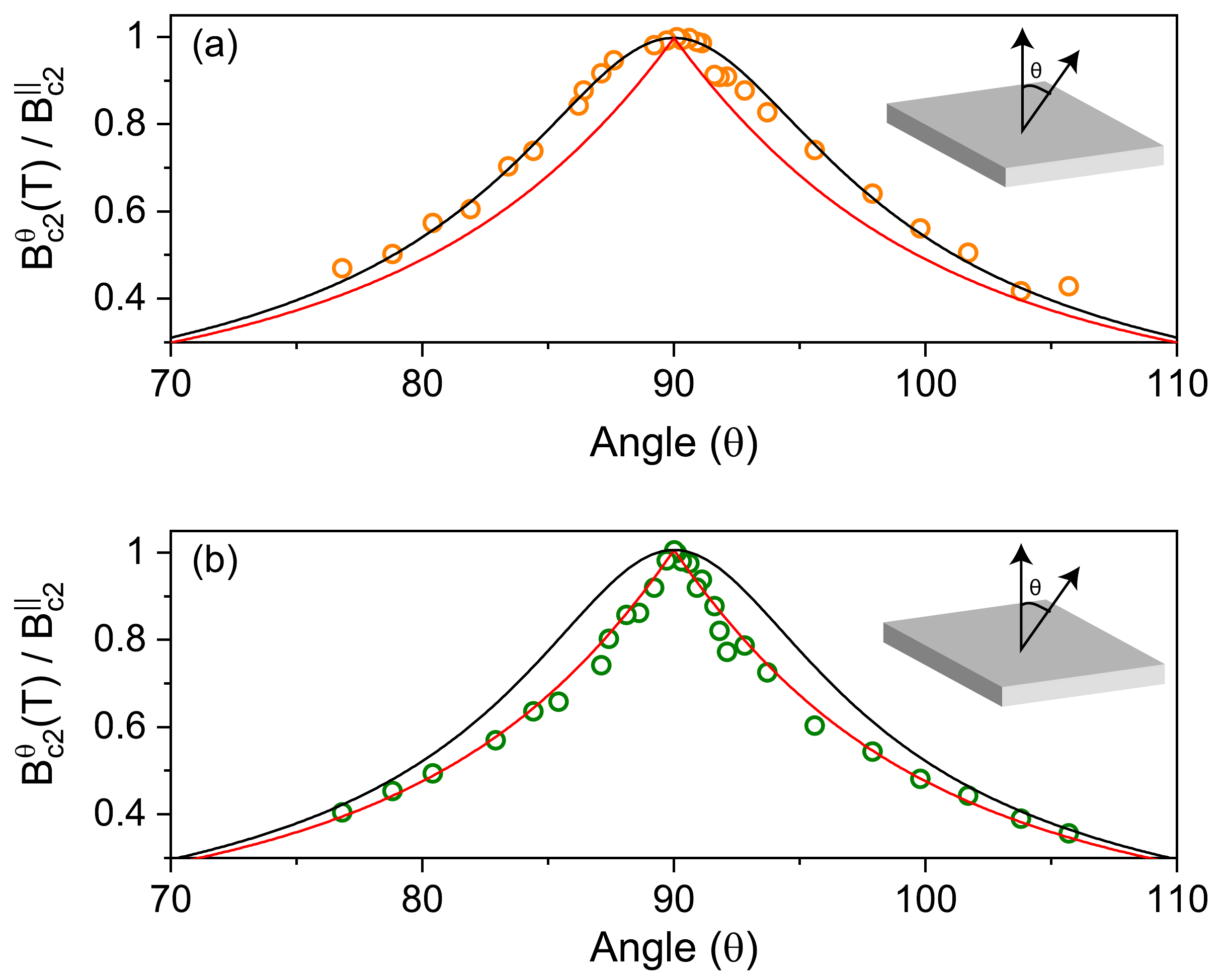}
		\caption{\small {Angular dependence of the $B_{c2}$ for (a) region-I and, (b) region-II. The red and black lines in both the panels are plots of  2D Tinkham model and 3D GL anisotropic mass model,  respectively. The size of data points in both the plots are larger than the error in the data. The insets show the direction of $B_{c2}\left(\theta\right)$.} \label{Fig.2}}  
		\end{center}
	\end{figure}

Fig.~\ref{Fig.2}(a) and Fig.~\ref{Fig.2}(b) show the angle (measured with respect to the out-of-plane direction of the device) dependence of $B_{c2}$ normalized by the in-plane $B_{c2}$ at $T=0.9T_c$ for region II and region I, respectively. The angular dependence of $B_{c2}(\theta)$ for the region II is well described by the 2D Tinkham model ($\left(B_{c2}\left(\theta\right)|\cos{\theta}|/B_{c2}^\perp\right)+\left(B_{c2}\left(\theta\right)\sin{\theta}/B_{c2}^{||}\right)^2=1$)~\cite{tinkham1963effect}. On the other hand, $B_{c2}\left(\theta\right)$ in the \ch{NbSe2} region follows the 3D Ginzburg-Landau model ($\left(B_{c2}\left(\theta\right)\cos{\theta}/B_{c2}^\perp\right)^2+\left(B_{c2}\left(\theta\right)\sin{\theta}/B_{c2}^{||}\right)^2=1$)~\cite{tinkham2004introduction} establishing that in contrast to the 3D superconductivity of the pure \ch{NbSe2} in region-I, the superconductivity in region-II is of 2D nature. This is the central point of this letter. 

Having established that the SC in region-II is 2-dimensional, we estimate the Berezinskii-Kosterlitz-Thouless transition temperature, $T_{BKT}$ for region-II from both non-liner current-voltage characteristics and from the $R-T$ characteristics~\cite{minnhagen1987two,finotello1985universality}) to be $6.14$~K(see Appendix~\ref{D} for details).

Fig.~\ref{Fig.3}(a), and Fig.\ref{Fig.3}(b) show the normalized magnetoresistance as a function of a magnetic field applied in-plane for the region-II and region-I, respectively. The in-plane critical field ($B^{||}_{c2}$) are plotted versus the reduced  temperature, $T/T_c$ in Fig.~\ref{Fig.3}(c). Note that $B^{||}_{c2}$ for the region-II is significantly larger than that of the pristine \ch{NbSe2} in region-I. The data from region-II fits well with the 2D GL equation $B_{c2}=B_{c2}\left(0\right)\sqrt{1-T/T_c}$ ~\cite{tinkham2004introduction} giving a $B^{||}_{c2}\left(0\right)\sim 36.5$~T. This is well beyond the Pauli Paramagnetic limit, $B_P$ ($=\sim1.86T_c$) $\sim10.5$~T for the system. For the region-I the data, as expected for  \ch{NbSe2} flake thicker than the superconducting coherence length, fits with the 3D formula, $B_{c2}=B_{c2}\left(0\right)\left(1-T/T_c\right)$~\cite{tinkham2004introduction}. The fact that the SC phase in region-II has a dimension d=2 (in contrast to d=3 for \ch{NbSe2} in region-I)  is the central result of this letter.  
		
		\begin{figure}[t]
			\begin{center}
				\includegraphics[width=\columnwidth]{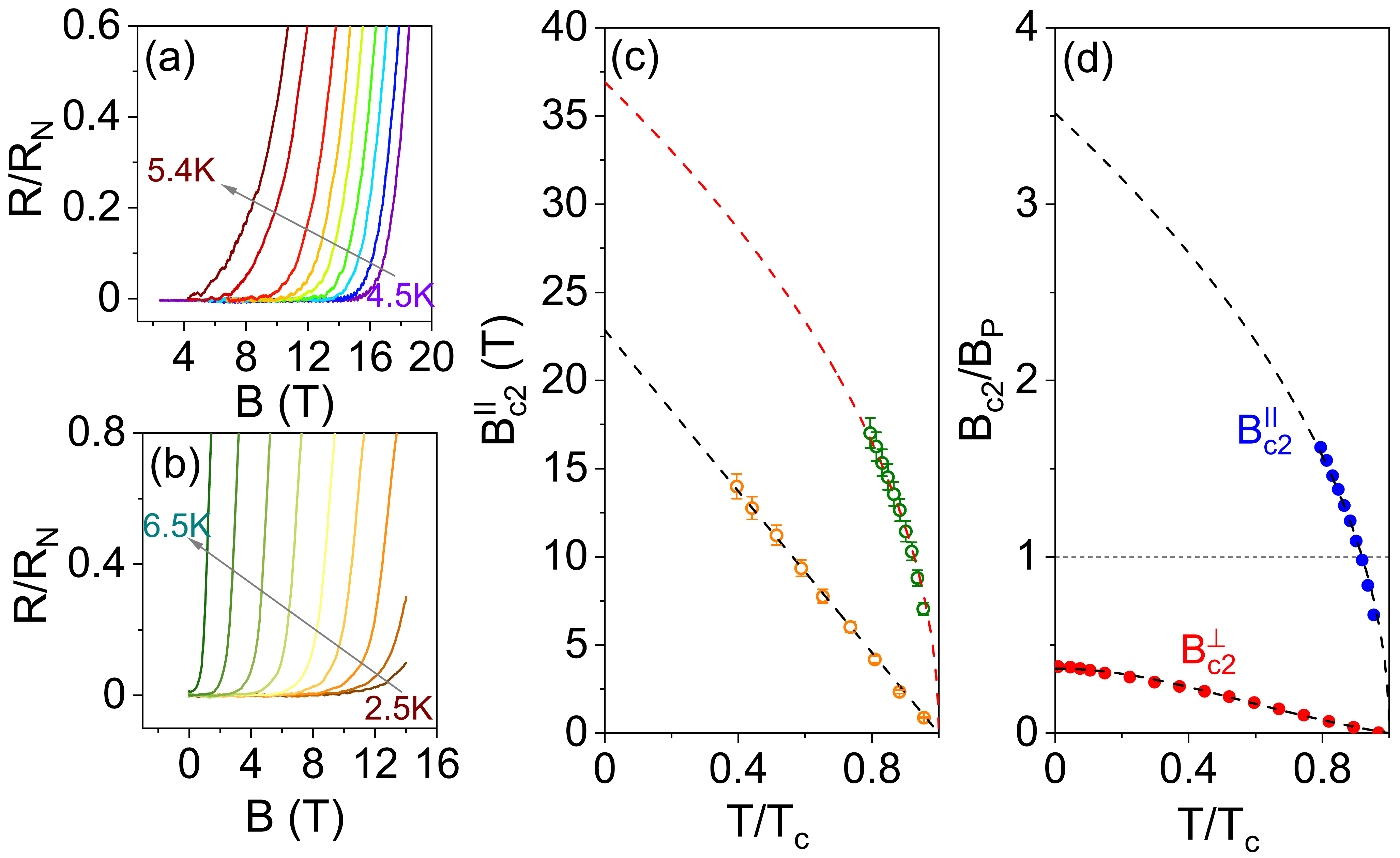}
				\caption{\small {Plot of magnetoresistance of the (a)  region-I and (b)  region-II (normalized in both cases by the normal state resistance) for in-plane magnetic field. (c) Plot of $B_{c2}^{||}$ versus $T/T_c$ for  region-I (green open circles) and region-II (orange open circles)  with the 2D GL fit (dashed red line) for region-II and 3D fit (dashed black line) for region-I. (d) Plot of $B_{c2}/B_P$ versus $T/T_c$ for both out of plane (red circle) and in-plane (blue circle) magnetic field direction for region-II -- the dashed lines show the fits to the data. } \label{Fig.3}}   
			\end{center}
		\end{figure}

Fig.~\ref{Fig.3}(d) shows the plots of both $B^{||}_{c2}$ and $B^{\perp}_{c2}$ (normalized by $B_P$) versus the reduced temperature for region II. Here $B_P$ ($= 1.86T_c$)$\sim10.5$~T is the Pauli paramagnetic limit for the SL-\ch{MoS2}/\ch{NbSe2} heterostructure.  As can be seen, $B^{||}_{c2}\left(0\right)$ is $\sim3.55$ times above the Pauli limit. This is a direct indication of Ising superconductivity. The spin-splitting energy, $\Delta_{SO} \approx 2\mu_B B^{||}_{c2}\left(0\right)^2/B_P$ ~\cite{sigrist2009introduction,youn2012role,xi2016ising} estimated from the parallel-field magnetoresistance data was $\sim$~15meV which matches quite well with the theoretical estimation of Zeeman type SOI, $\Delta_{SO}$ $\sim13$~meV in 2-dimensional TMD~\cite{saito2016superconductivity}.
			
\begin{figure}[t]
			\begin{center}
				\includegraphics[width=\columnwidth]{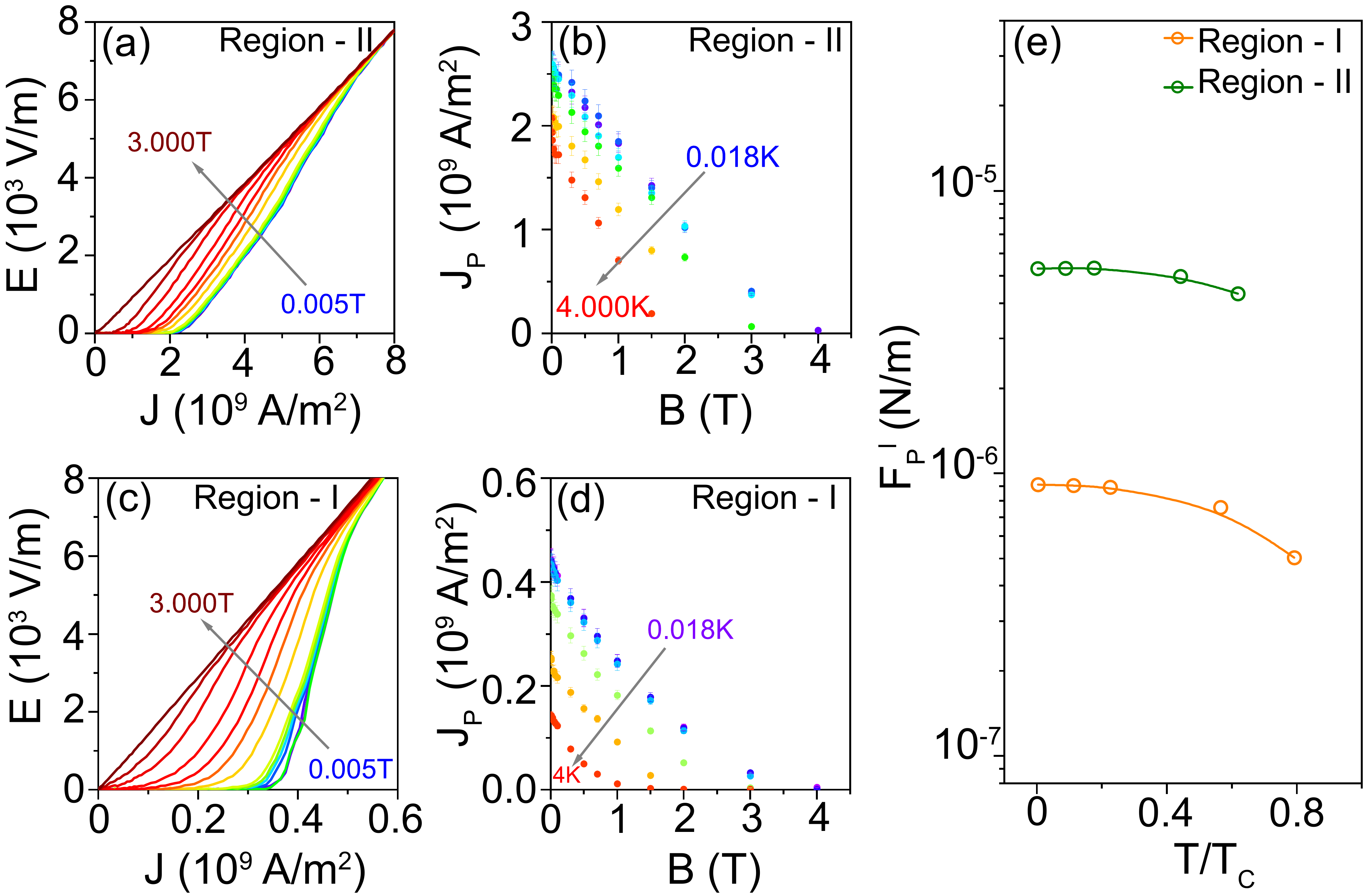}
				\caption{\small {$E-J$ characteristics at different $B^\perp$ measured at $T=2.5$~K for the (a) region-II, and (b) region-I. Depinning current density $J_p$ as a function of $B^{\perp}$ at $2.5$~K for the (c) region-II, and (d) region-I. (e) Plots of the pinning force per unit length of the vortex, $F_P^l$ versus the reduced temperature $T/T_c$ region-I (orange filled circles) and region-II (olive filed circles). The lines are guided to the eye.}\label{Fig.4}}  
			\end{center}
		\end{figure}

To further distinguish between transport in the two regimes, we probed the vortex dynamics of the system through measurements of DC non-linear current-voltage characteristics in the presence of an out-of-plane magnetic field, $B^\perp$. Fig.~\ref{Fig.4}(a) and (c) show the $E$-$J$ characteristics for the two regions of the device measured at $2.5$~K for different values of $B^\perp$ (the data for another device are presented Appendix~\ref{F}). For a disordered superconductor in the flux-flow regime, the charge current density is given by $E=\ \rho_{ff}\left(J-J_p\right)$~\cite{orlando1991foundation}, where $E$ is the electric field between the voltage probes, $\rho_{ff}$ is the flux flow resistivity, and $J_p$ the depinning current density. A comparison of Fig.~\ref{Fig.4}(b) and (d) shows that $J_P$ for the region-II is much higher than that in region-I. The pinning force per unit length of the vortex, extracted using the equation $F_P^l=\ J_p (h/2e)$~\cite{blatter1994vortices}, is plotted in Fig.~\ref{Fig.4}(e) at a few representative values of  $T/T_c$. The value of $F_P^l$ for region-II is an order of magnitude higher than that of region-I. This is expected given the significantly higher defect levels in SL-\ch{MoS2}~\cite{sarkar2019probing} as compared to that in \ch{NbSe2}. This establishes that in region-II of the device, the presence of \ch{MoS2} has a major effect on the transport of supercurrent through the underlying \ch{NbSe2}.  
		
%

To recapitulate our principal results, we observe superconductivity at the heterojunction of single-layer \ch{MoS2} and few-layer \ch{NbSe2}($\sim15$~nm) which differs from that of \ch{NbSe2} in several different aspects -- the primary being its 2-dimensional Ising nature.  We turn now to the discussion of the possible origin of this Ising SC. 

A plausible mechanism is that the SL-\ch{MoS2} can have cracks facilitating current to flow through it into the \ch{NbSe2},  This artefact was ruled out from AFM topography mappings on multiple devices, which did not show any structural damage on the SL-\ch{MoS2} upon transfer on the pre-patterned hBN probes (See Appendix~\ref{C}). An alternate proposition is that the transport in region-II is solely through the underlying \ch{NbSe2} whose superconducting properties change from that of 3D to 2D-like due to the presence of the single-layer~\ch{MoS2}. We note that further experimental and theoretical studies are necessary to ascertain the mechanism of superconductivity in this hybrid system.

\begin{figure}[t]
	\begin{center}
		\includegraphics[width=\columnwidth]{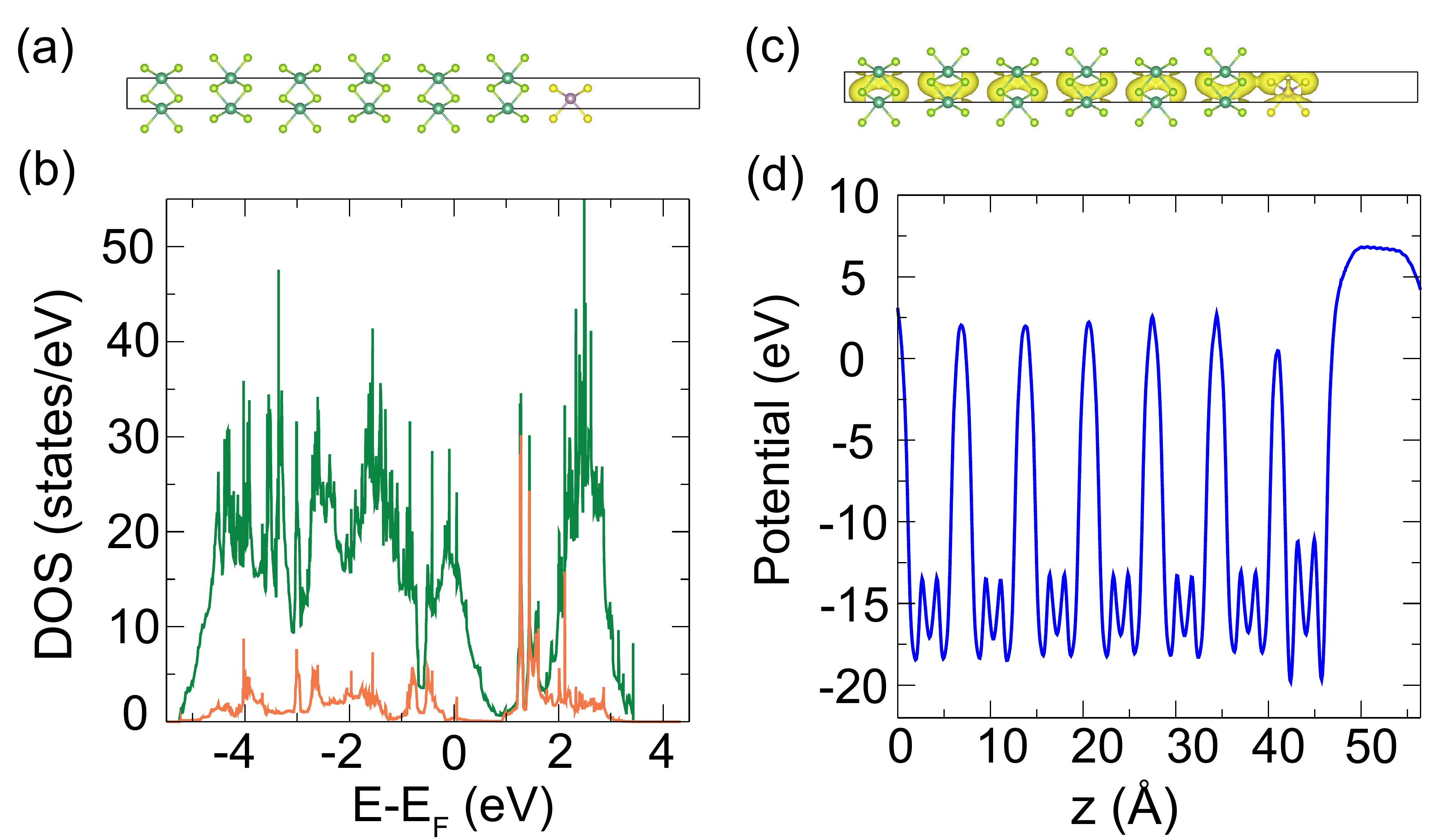}
		\caption{\small {(a) Illustration of the heterojunction with $6$ layers of \ch{NbSe2} and $1$ layer \ch{MoS2}. (b) The total density of states (olive) and density of states projected on the \ch{MoS2} layer (orange) of the system. (c) Charge density isosurfaces for the heterojunction.   (d) The average local potential along the stacking direction, $z$.}\label{Fig.5}}  
	\end{center}
\end{figure}

In an attempt to understand the effect of the presence of the single-layer \ch{MoS2} on the superconductivity of the underlying  \ch{NbSe2}, we carried out first-principles density functional theory calculations (see Appendix~\ref{H} for technical details). We constructed a heterojunction comprising of $6$ layers of \ch{NbSe2} and $1$ layer of \ch{MoS2}, as shown in Fig.~\ref{Fig.5}(a). The resulting density of states is presented in Fig.~\ref{Fig.5}(b). The total density of states (olive curve) shows that the system is metallic, with many states around the Fermi level. We further calculate the density of states projected on the \ch{MoS2} layer (orange curve). Notably, the \ch{MoS2} layer has also lost its semiconducting character when in proximity to the \ch{NbSe2}, as illustrated by the finite density of states around the Fermi level on the \ch{MoS2} layer. To further corroborate our findings, we present a plot of the charge density of the heterojunction in Fig.~\ref{Fig.5}(c). The isosurfaces show substantial charge density at the various layers. Most notably, we find that the charge densities of the SL-\ch{MoS2} layer and the \ch{NbSe2} layer in contact with it are overlapping, confirming the hybridization between the two. This hybridization may result in the change in superconducting properties of \ch{NbSe2} from 3D to 2D-like, as we observed. From the Hall data in Appendix~\ref{E} Fig.~\ref{Fig.S5} the overlap of charge density is evident. We can also notice that the adding the SL-\ch{MoS2} in region-II varies the density from that of the pristine \ch{NbSe2}. The implication of this is that the coherence length of region-II becomes larger than the typical $\sim9$~nm value of pristine \ch{NbSe2} which consequently lead to the observation of 2D superconductivity in heterostructure.
		
An interesting observation is that the charge carriers in region-II are holes -- this is in accordance with previous observations~\cite{guan2017optimizing}. Our observation that region-II becomes a hole-doped 2-dimensional Ising SC is exciting because of the prospect that it can host a topologically non-trivial superconducting phase~\cite{triola2016general,hsu2017topological} and SC with finite-momentum-pairing per the predictions by Fulde and Ferrell~\cite{PhysRev.135.A550} and by Larkin and Ovchinnikov~\cite{LO}. This makes our system unique; as in all earlier reports, SC in TMD was achieved through electron doping~\cite{ye2012superconducting}. 

\section{Conclusion}		
In summary, we have observed a transition from 3-dimensional to 2-dimensional superconductivity in heterostructure of few-layer \ch{NbSe2}($\sim15$~nm) and SL-\ch{MoS2}. Magnetotransport measurements establish that the superconductivity in the underlying \ch{NbSe2}, albeit physically in the 3-dimensional limit,  has a strong Ising pairing. The stability of the superconducting phase and the system being in the hole-doped region makes our device structure unique and opens up the scope of observing topological superconductivity in van der Waals systems. 			

\begin{acknowledgements}
	
	The authors acknowledge device fabrication facilities in NNFC, CeNSE, IISc. A.N. acknowledges support from the startup grant (SG/MHRD-19-0001) of the Indian Institute of Science. A.B acknowledges funding from SERB (HRR/2015/000017), DST (DST/SJF/PSA-01/2016-17), and IISc.
	
\end{acknowledgements}	

\cleardoublepage

\section*{Supplementary Materials}
\renewcommand{\thesection}{S\arabic{section}}
\setcounter{section}{0}

\renewcommand{\thefigure}{S\arabic{figure}}
\setcounter{figure}{0}
	\section{Sunken electrical probes on hBN substrate \label{A} }

	\begin{figure}[h]
		\begin{center}
			\includegraphics[width=\columnwidth]{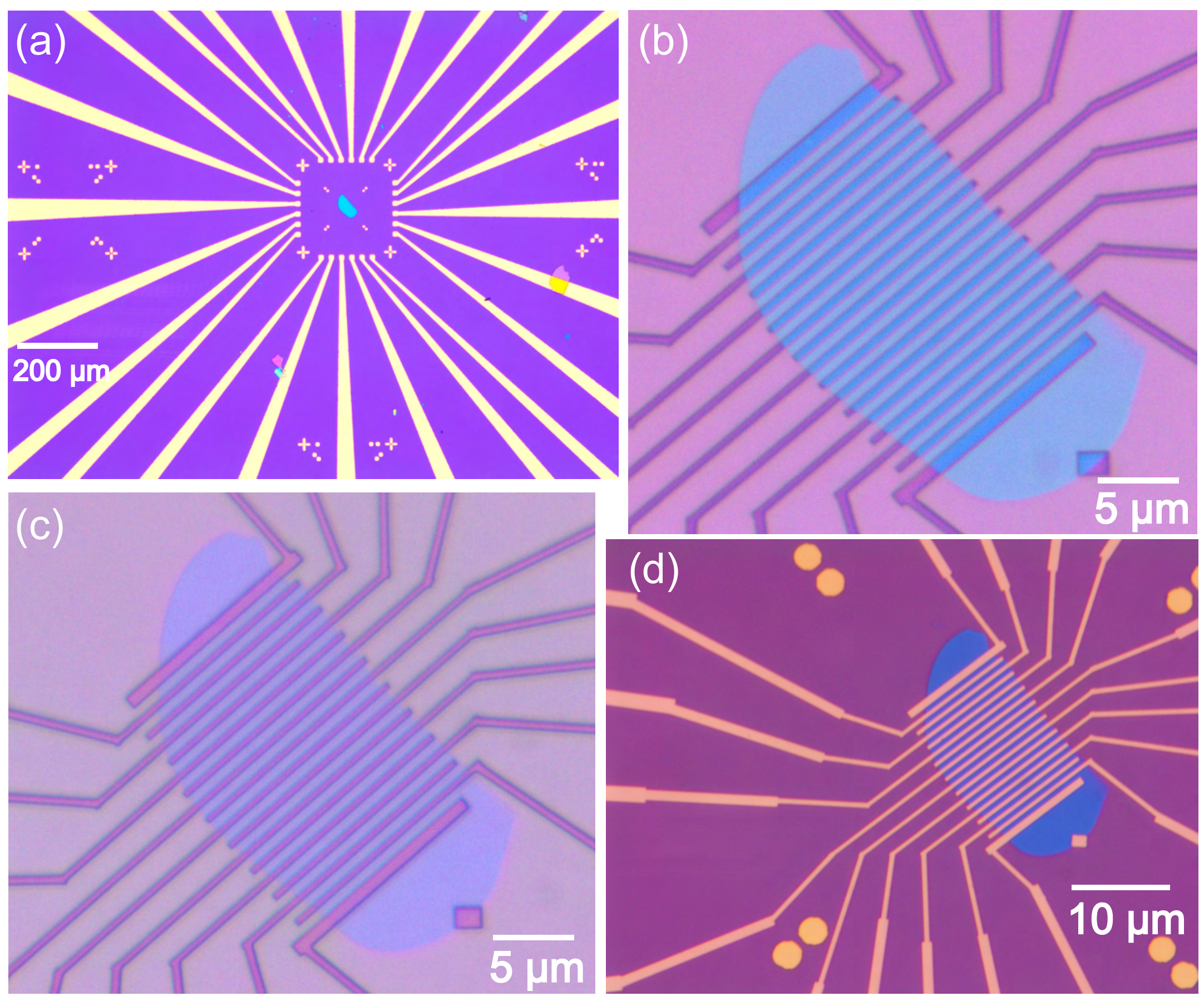}
			\caption{\small {Optical image at different stages of fabrication process of a \ch{hBN} sunken \ch{Au} probe (a)$\sim25$~nm \ch{hBN} flake transferred on pre-patterned substrate. (b) Electron-beam lithography patterns for contact on the \ch{hBN} flake before dry etching. (c) The \ch{hBN} flake after the dry etching process (d) Final substrate after metal deposition.}\label{Fig.S1}}   
		\end{center}
	\end{figure}

	\begin{figure}[t]
		\begin{center}
			\includegraphics[width=\columnwidth]{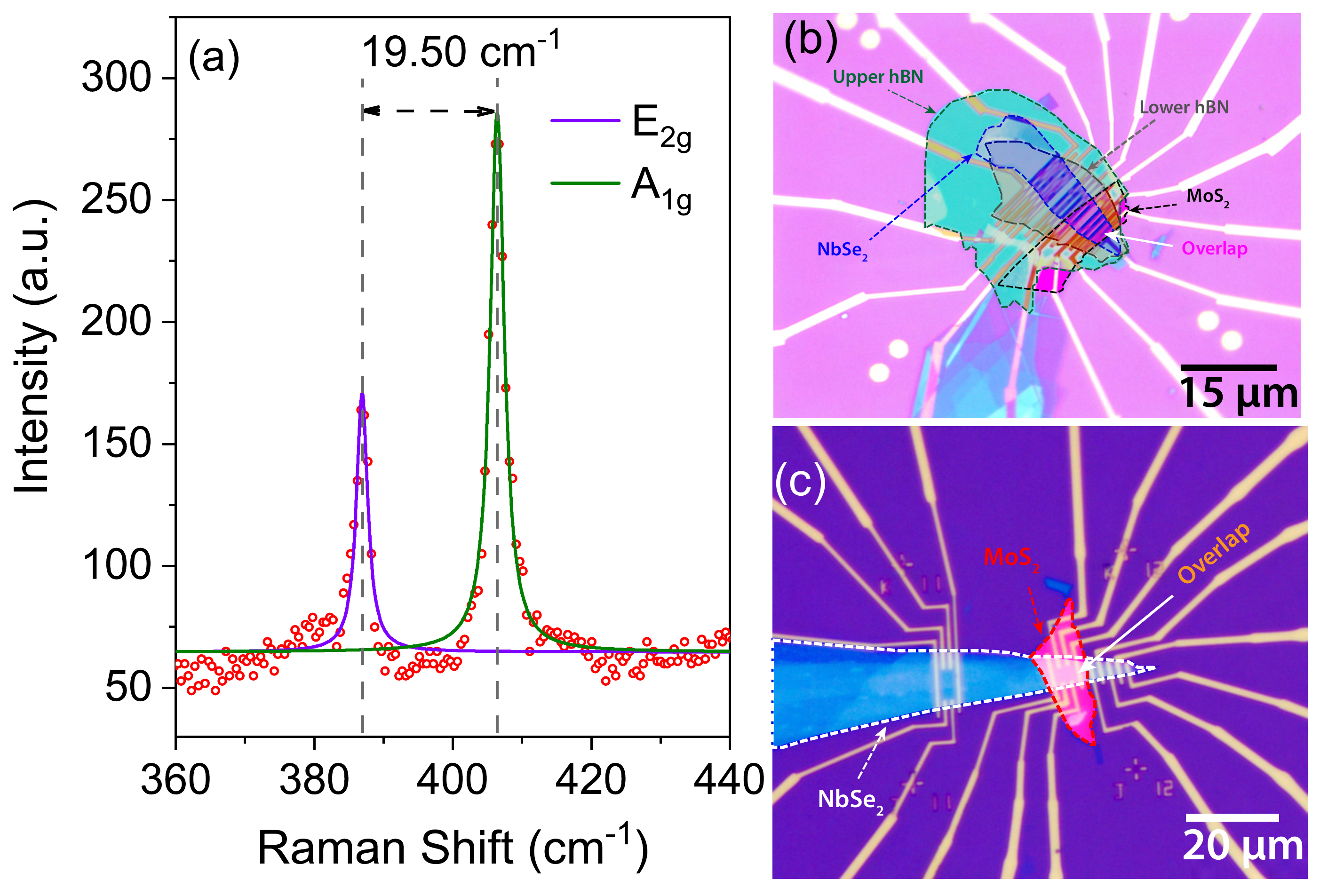}
			\caption{\small {(a) Room temperature Raman spectra of \ch{MoS2} (red open circles) along with the Lorentzian fits for the $E_{2g}$ peak (violet line) and $A_{1g}$ peak (pink line). The difference in Raman Shift between the two peaks is $19.50$~cm$^{-1}$  indicating the \ch{MoS2} to be a single-layer. False color optical image of few other measured devices in two different probe geometry (b) Linear probe (c) Hall measurement device for estimation of carrier no density discussed in Appendix~\ref{E}.  } \label{Fig.S2}}   
		\end{center}
	\end{figure}

	One disadvantage of dealing with transition metal dichalcogenide (TMD) materials, especially \ch{NbSe2},  is their low stability in ambient conditions \cite{xi2016gate}. Keeping this fact in mind we went for an encapsulated device structure (as described earlier). The flakes were transferred on pre-fabricated \ch{hBN} sunken \ch{Au} probes to avoid degradation due to exposure to moisture as chemicals such as e-beam resist, organic solvents etc. The `sunken' probes also help avoid strain (usually present in such structures~\cite{kundu2017quantum}) to the thin flakes of \ch{MoS2} which may result in a `tear' that can electrically short it with the \ch{NbSe2} above it. So for this purpose we first fabricated a pre-patterned substrate where big electrical contact pads of dimensions $\sim 200~\mu$m x $200~\mu$m were initially patterned using photo-lithography technique on a $\sim 2$~mm x $2$~mm piece of a \ch{Si}/\ch{SiO2} substrate. The \ch{SiO2} in the exposed region of the pattern were then etched by $\sim 45$~nm using a Buffered \ch{HF} solution with a subsequent step of deposition of \ch{Cr}-\ch{Au} of thickness $5$-$55$~nm respectively creating big pads  of an effective height of $15$~nm with respect to the substrate surface. \ch{hBN} flakes were procured using micromechanical exfoliation with standard scotch-tape method \cite{novoselov2004electric} on commercial PDMS gel stamp(Gel-Pak,PF-$3$-X$4$). Flakes of thickness $\sim 25$~nm were roughly selected using optical microscope contrast. Selected flakes were then transferred using a micro-manipulator and an optical microscope in the middle section of a pre-patterned probes on \ch{Si}/\ch{SiO2} substrate as shown in Fig.~\ref{Fig.S1}(a). After the transfer of \ch{hBN} onto the substrate the thickness of the flake were confirmed using AFM technique and substrates having \ch{hBN} flakes of thickness $\sim 25\pm 2$~nm were only selected for further process. Subsequently patterns of electrical contacts on \ch{hBN} as well as the connecting lines between \ch{hBN} and the pre-deposited big pads were defined using e-beam lithography technique (Fig.~\ref{Fig.S1}(b)) with a follow up step of selective etching of \ch{hBN} using a proportionate mixture of \ch{CHF3} and \ch{O2} in a Reactive Ion Etching system(RIE-\ch{Cl}) to etch off the \ch{hBN} in the exposed part of the flake along the contact lines as shown in Fig.~\ref{Fig.S1}(c). As the final step \ch{Cr}-\ch{Au} of thickness $5$-$22$~nm were then deposited giving an almost planar contact surface on \ch{hBN} (Fig.~\ref{Fig.S1}(d)). The advantage of having only $15$~nm of effective height of big pads is that the connecting lines of total thickness of $\sim 27$~nm would `climb' on top of it upon deposition ensuring continuous electrical connection. As a final step the prepared probes were vacuum annealed at $250^{\circ}$C for $3$ hours to get rid of any residue on the \ch{hBN} surface before further process. 
	
		\section{Details of  device fabrication \label{B}}

	Single-layer flakes of \ch{MoS2} were obtained from high-quality single crystal by mechanical exfoliation~\cite{novoselov2004electric} on PDMS gel stamp. The thickness of the flakes was verified through room-temperature Raman spectroscopy~\cite{wang2012electronics}. Fig.~\ref{Fig.S2}(a) is the Raman spectra of one such \ch{MoS2} flake – shows the presence of $E_{2g}$ and $A_{1g}$ peaks with a peak difference of $\sim19.5$~cm$^{-1}$ in Raman shift confirming the flake to be a single-layer~\cite{wang2012electronics}. The \ch{MoS2} flake was transferred partially on the pre-patterned Au contact probes on hBN  (see Appendix~\ref{A}) using a dry transfer technique~\cite{yang2014multilayer} with the help of a high precision electrical micromanipulator and a digital camera fitted with an optical microscope. AFM measurements were carried out to rule out any cracks in the \ch{MoS2} layer (see Appendix~\ref{C}).

	\begin{figure}[t]
		\begin{center}
			\includegraphics[width=\columnwidth]{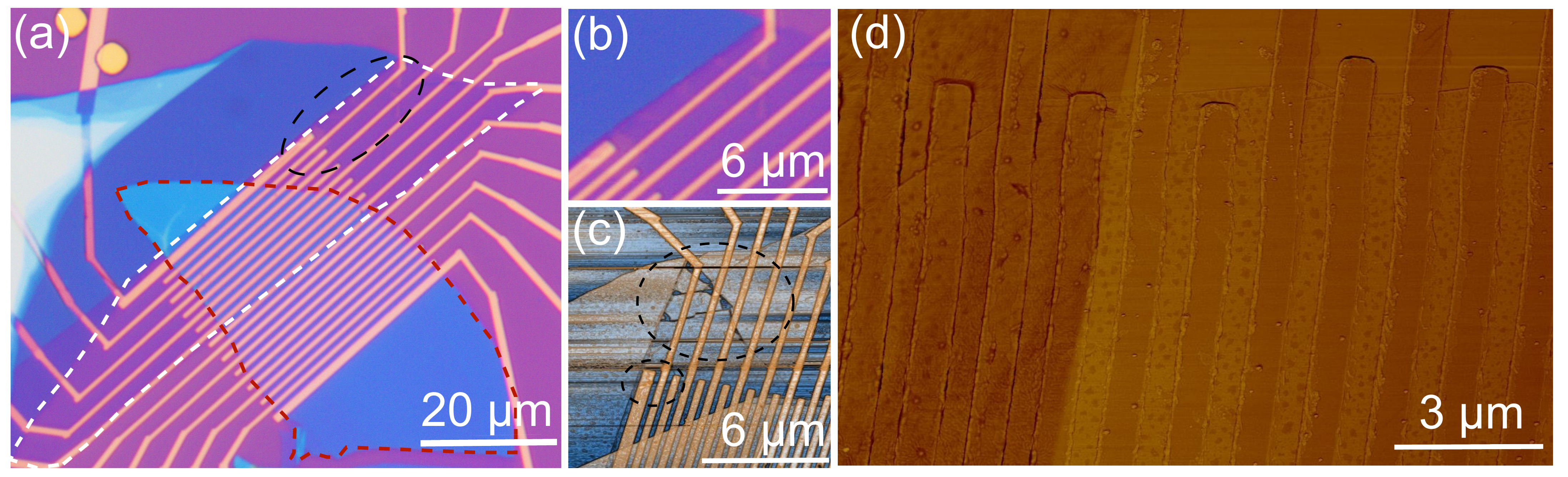}
			\caption{\small{(a) Optical image of monolayer of \ch{MoS2} transferred on hBN sunken gold probes with false color dashed line showing the boundary of the \ch{MoS2 }(white) and hBN (red). The black dashed  circle marks a region with cracks in the \ch{MoS2} layer. (b) Zoomed in image of the area marked by circle in Fig.~\ref{Fig.S3}(a) showing cracks in \ch{MoS2} (c) AFM image of the system in enhanced coloring. (d) Zoomed in AFM image of the section on probe showing \ch{MoS2} on probe (darker shade) and bare probes (lighter shade).}\label{Fig.S3}}
		\end{center}    
	\end{figure}

	\ch{NbSe2} flakes were exfoliated in similar fashion in the inert environment of a glove-box having \ch{O2} and \ch{H2O} concentrations $<0.5$~ppm. The thickness of the flake was estimated to be  $\sim15$~nm from optical contrast. The \ch{NbSe2} flake was then transferred very precisely using the remotely controlled micromanipulator onto the probes such that half of the \ch{NbSe2} flake lay on the exposed \ch{Au} probes, whereas the other half got transferred on the single-layer \ch{MoS2}. We refer to the region having only \ch{NbSe2} as ‘region-I’ while the region with \ch{MoS2} on \ch{NbSe2} is referred to as ‘region-II’. Finally, an hBN flake of thickness $\sim30$~nm was transferred to the device to avoid environmental degradation of the heterostructure. As a final step, the device was  vacuum annealed at ${200}^\circ$~C to improve the coupling between the \ch{MoS2} and \ch{NbSe2} layers~\cite{sarkar2019evolution}.         
	
	\section{Verifying the topography of \ch{MoS2} transferred on probes \label{C}}

	To choose regions of defect-free single-layer \ch{MoS2}, AFM measurements were carried out after transfer on the Au-probes. Fig.~\ref{Fig.S3} shows an example of both the optical (Fig.~\ref{Fig.S3}(a)) and AFM (Fig.~\ref{Fig.S3}(c)) images. The regions having cracks on \ch{MoS2} (marked with blacked circles) as shown in Fig.~\ref{Fig.S3}(b) gives a contrast difference both in optical as well as AFM which makes it easier to identify them. These areas were stringently avoided in making the final device. The zoomed in image of the single-layer \ch{MoS2} on the Au-probes in Fig.~\ref{Fig.S3}(d) also shows the continuity of flakes. The few bubbles which are present in Fig.~\ref{Fig.S3}(d) were eliminated by the vacuum annealing after final assembly of the entire device.
	
	\section{Establishing BKT Physics \label{D}}
	
	\begin{figure}[t]
		\begin{center}
			\includegraphics[width=\columnwidth]{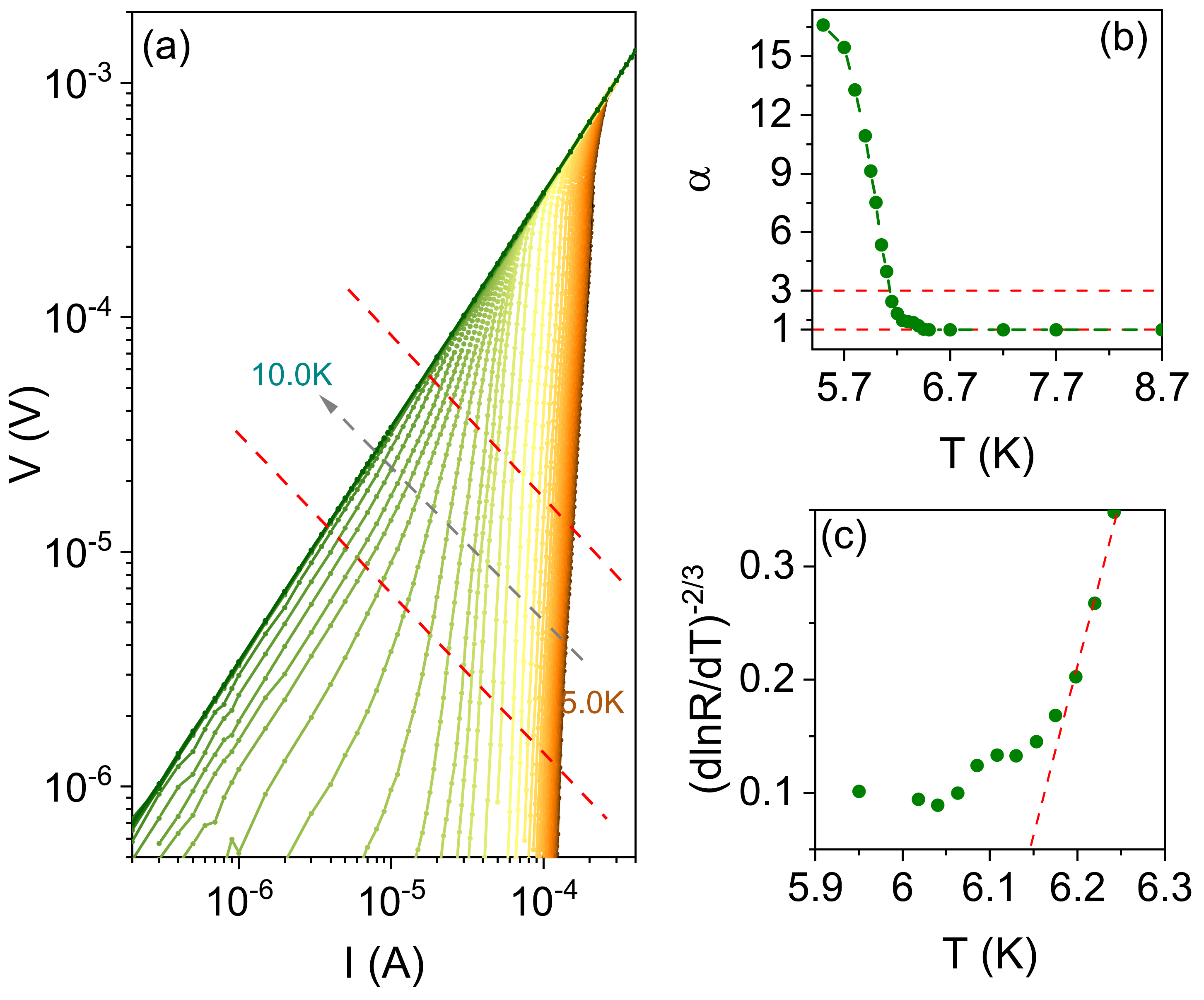}
			\caption{\small{(a) Zero magnetic field current-voltage characteristics of the region-II with varying temperature ranging from 5.0~K to 14.0~K. The red dashed line represents the range of current within which linear fit was done for each curve. (b) Plot of $\alpha$ as a function of temperature, $T$ to evaluate $T_{BKT}$ for  region-II. (c) Plot of $(dlnR/dT)^{-{2/3}}$ as a function of temperature, $T$ for the \ch{MoS2} where the intersect of the black dashed line with the x axis gives $T_{BKT}$.}\label{Fig.S4}}
		\end{center}
	\end{figure}
	
	In the main text, we have established, through angular dependence of the critical field, that in region-II the superconductivity has an effective dimensionality of two. In this section we evaluate the Berezinskii-Kosterlitz-Thouless transition temperature, $T_{BKT}$ of region-II from the temperature dependent current-voltage characteristics (Fig.~\ref{Fig.S4}(a)). For a 2D superconductor, electric field induced unbinding of the the vortex-antivortex pairs give rise to non-linear current-voltage characteristics of the form $V \sim I^\alpha$. At $T=T_{BKT}$, the non-linearity exponent, $\alpha$ becomes $3$ as shown in Fig.~\ref{Fig.S4}(b)~\cite{minnhagen1987two,minnhagen1983two}. This analysis yields $T_{BKT}=6.13$~K.
	
	One can also evaluate $T_{BKT}$ from the $R$-$T$ characteristics using the relation $R = R_0\mathrm{exp}[-b_R/(T-T_{BKT})^{1/2}]$, where $b_R$ is quantity indicating the vortex-antivortex interaction strength~\cite{minnhagen1987two,ambegaokar1980dynamics,finotello1985universality}. The formula is valid over a very small range of above $T_{BKT}$ as within that range superconductivity is destroyed by thermal unbinding of vortex-antivortex pairs. Thus by plotting its reduced form $\left(dlnR/dT\right)^{-2/3} = \left(2/b_R\right)^{2/3}\left(T-T_{BKT}\right)$ we can deduce the $T_{BKT}$ from the x-axis intercept as shown in Fig.~\ref{Fig.S4}(c) which comes out to be $6.14$~K. The estimation of $T_{BKT}$ from both the analysis is very close to the value of $6.3$~K reported for ion-gated TMD in earlier results~\cite{lu2015evidence}. 
	
	\section{Hall measurements \label{E}}
	
	\begin{figure}[t]
		\begin{center}
			\includegraphics[width=\columnwidth]{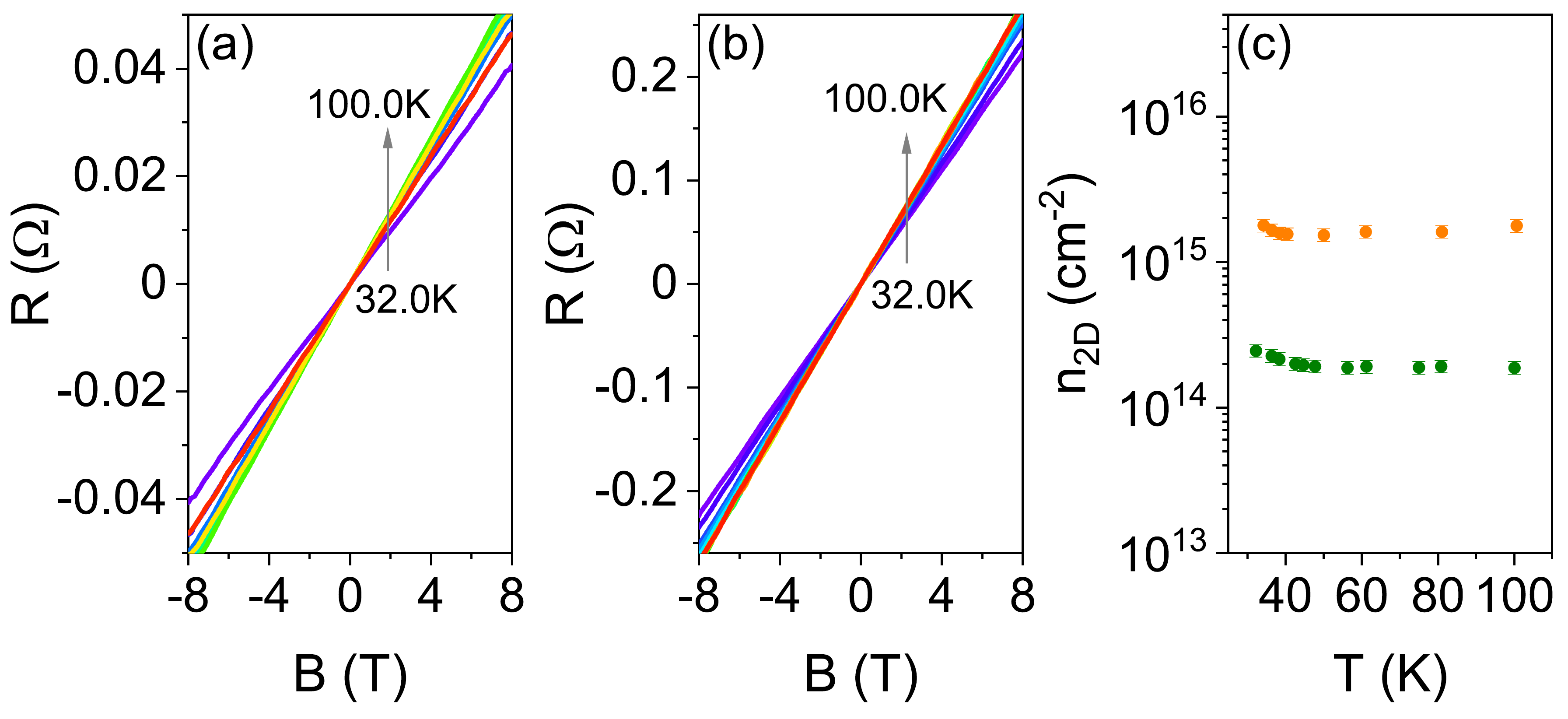}
			\caption{\small {Magnetic field ($B$) dependence of Hall resistance, $R_{xy}$ for (a) region-I over the temperature range $7.5$~K--$100$~K, and, (b) region-II over the temperature range $8$~K--$100$~K. (c) Plot of the charge carrier density, $n$ versus $T$ for both region-I (orange filled circles) and region-II (olive filled circles) showing similar carrier densities, i.e. holes.}\label{Fig.S5}}   
		\end{center}
	\end{figure}  
	
	Fig.~\ref{Fig.S5}(a) and Fig.~\ref{Fig.S5}(b) show respectively the plots of $R_{xy}$ versus $B$ measured for the region-I and region-II. The charge carrier densities calculated from these plots are shown Fig.~\ref{Fig.S5}(c). One can see that the estimated number densities in the region-II closely follows that of the pristine \ch{NbSe2} in region-I.

	\begin{figure}[t!]
		\begin{center}
			\includegraphics[width=\columnwidth]{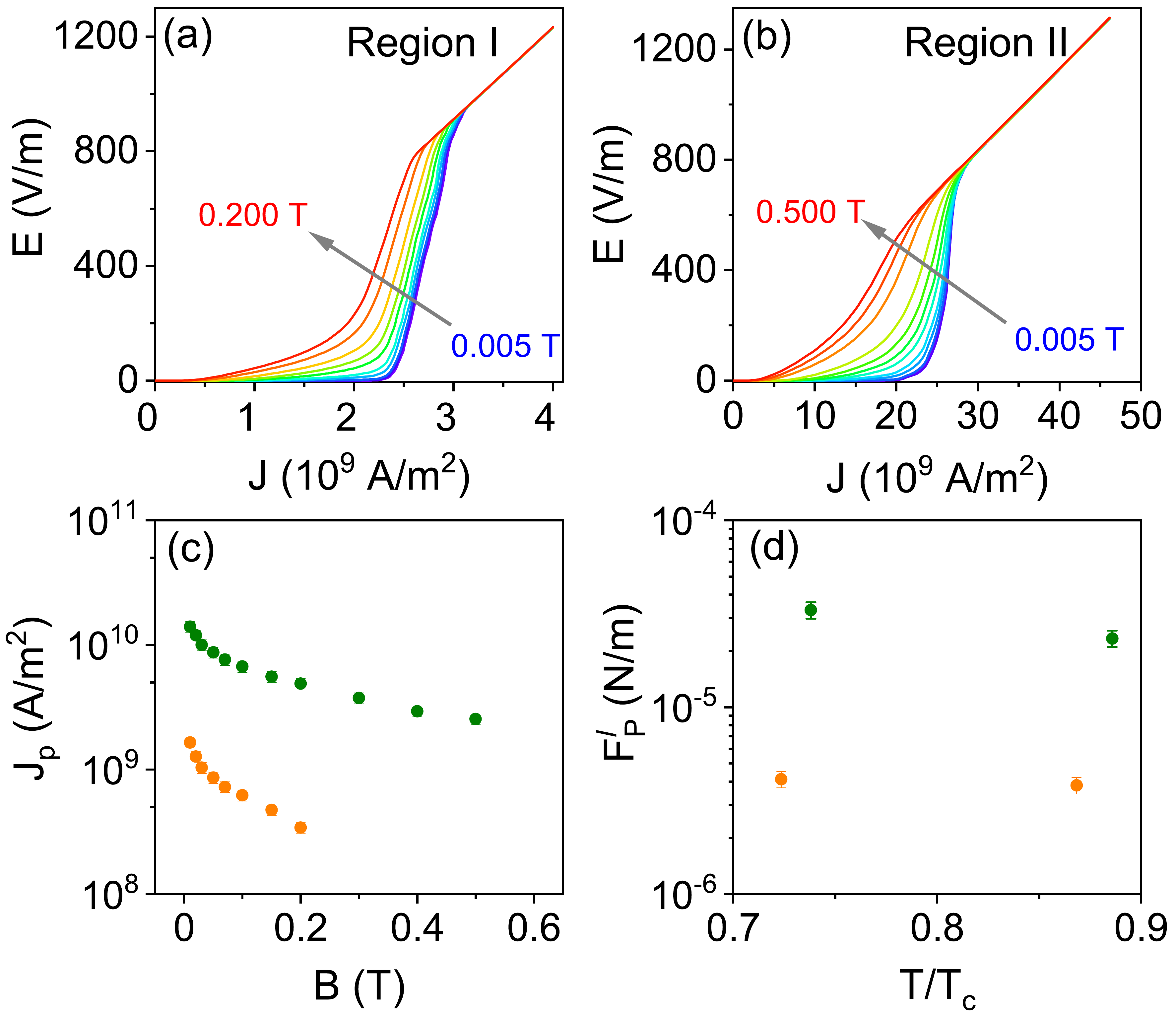}
			\caption{\small {$E-J$ characteristics at different perpendicular magnetic fields measured at $T=5$~K for the (a) region-I, and (b) region-II. (c) Depinning current density $J_p$ as a function of the applied perpendicular magnetic field $B$ at $5$~K for the  region-I (orange filled circles) and  region-I (green filed circles).(d) Plots of the pinning force per unit length of the vortex, $F_P^l$ as a function of reduced temperature $T/T_c$ for the  region-I (orange filled circles) and  region-II (green filled circles).}\label{Fig.S6}}  
		\end{center}
	\end{figure}
	
	\section{Depinning current,$J_p$ evaluation \label{G}}
	
	\begin{figure}[t]
		\begin{center}
			\includegraphics[width=0.7\columnwidth,height=0.5\textheight,keepaspectratio]{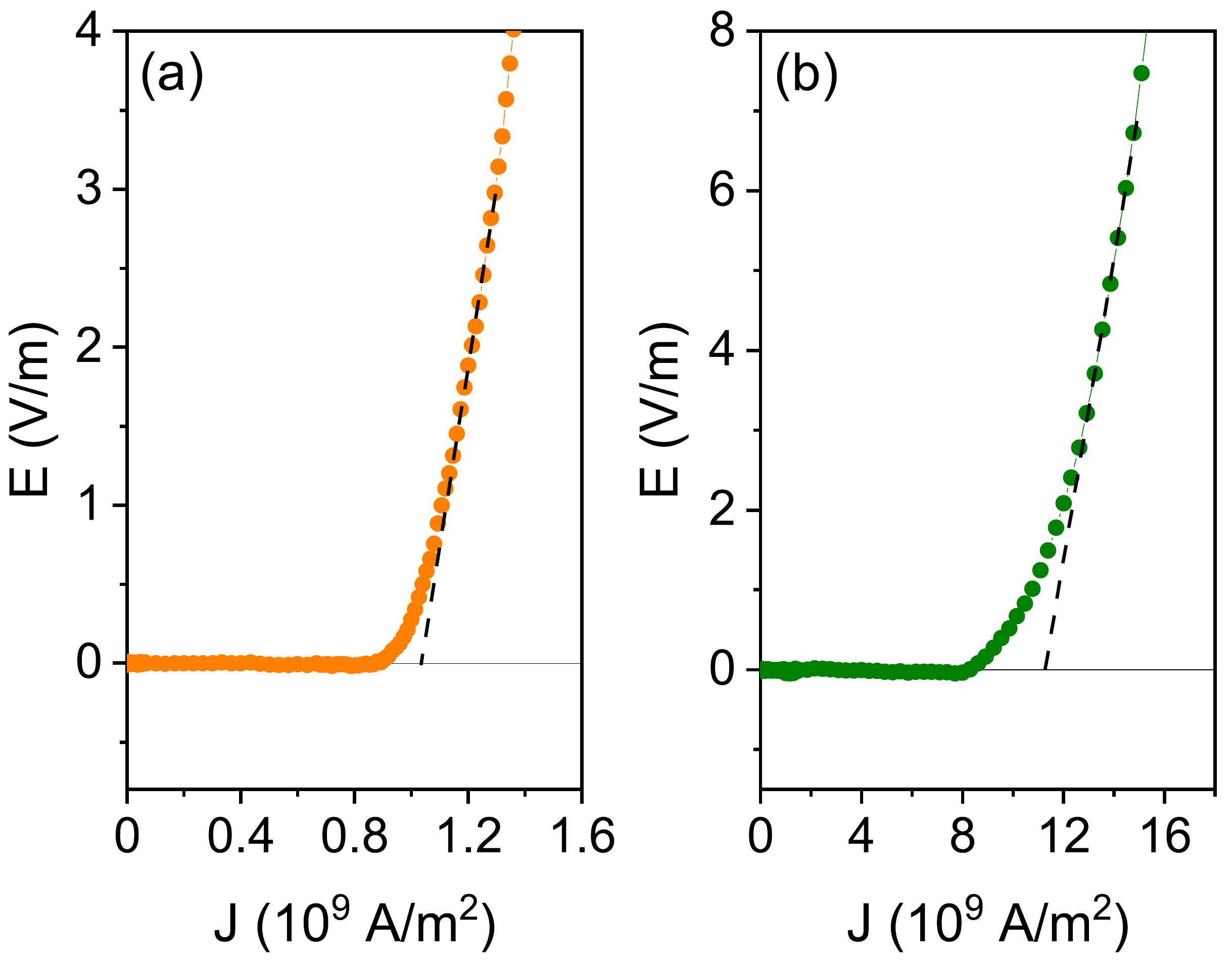}
			\caption{\small {$E-J$ characteristics at $T$= $5$~K and $B$=$0.03$~T for  (a) region-I, and (b) region-II. The dashed lines in both the plots are the linear fits - the intersect of this line with the $J$ axis gives an estimate of the depinning current density.}\label{Fig.S7}}   
		\end{center}
	\end{figure} 
	
	To obtain the depinning current, $J_P$ we have to consider the linear section of the $E-J$ curve (Fig.~\ref{Fig.S7}) near the onset of resistance as following the definition, for $J>J_P$ the system enters into the flux flow regime for which the resistance i.e. $R_{ff}$ becomes independent of current~\cite{orlando1991foundation}. For that purpose we select a section of the curve just above the $E$ value $1$~V/m and perform a linear fit in that region. We then extract the value of $J_P$ from the intersect of the best fit to the $J$ axis~\cite{kundu2019effect}, as shown in Fig.~\ref{Fig.S7}.

	\section{Ab-initio calculation \label{H}}
	
	Density functional theory computations were performed using the VASP code~\cite{kresse1996efficiency, PhysRevB.54.11169}. The Perdew-Burke-Ernzerhof approximation to the exchange-correlation functional, including the van der Waals correction, was employed~\cite{perdew1996generalized}. A plane wave cutoff of $300$eV was used, and the Brillouin zone was sampled using a $9\times9\times1$ $\Gamma -$ centered $k-$point mesh. All atoms were relaxed until the forces were less than $0.01$eV$ /$\AA. A vacuum of $10$~\AA{} was used to avoid spurious
    interaction between periodic images.

\end{document}